\documentclass[12pt]{article}
\usepackage{amssymb,tabularx,multirow,amsmath,natbib, epsfig, threeparttable, amstext, subfigure,xcolor,bbm, hyperref,wrapfig,graphicx,color,booktabs,array,makecell}

\usepackage{authblk}

\newcolumntype{P}[1]{>{\centering\arraybackslash}p{#1}}
\newcolumntype{L}[1]{>{\raggedright\arraybackslash}p{#1}}
\newcolumntype{R}[1]{>{\raggedleft\arraybackslash}p{#1}}

\usepackage[margin=1.0in]{geometry}
\usepackage[font=small]{caption}

\title{Commanding the Foul Shot: A New Ensemble of Free Throw Metrics}

\author[1,2,*]{Jake McGrath}
\author[3]{Amanda Glazer}
\author[2]{Vanna Bushong}
\author[2]{Michelle Nguyen}
\author[2]{Kirk Goldsberry}

\affil[1]{Department of Physics, University of Texas at Austin}
\affil[2]{Business of Sports Institute, McCombs School of Business, University of Texas at Austin}
\affil[3]{Department of Statistics and Data Sciences, University of Texas at Austin}
\affil[*]{\texttt{jmcgrath@utexas.edu}}

\date{}

\begin{document}

\maketitle

\begin{abstract}
    With the NBA’s adoption of in-game limb tracking in 2023, Sony’s Hawk-Eye system now captures high-resolution, 3D poses of players and the ball 60 times per second. Linking these data to key events opens a new era in NBA analytics. Here, we leverage a large dataset of 21,964 shot attempts from 72 NBA players to introduce a novel ensemble of metrics for evaluating free-throw shooting. Inspired by baseball analytics, we introduce command, which quantifies the quality of a free throw by measuring a shooter’s accuracy and precision near the basket’s bullseye. This metric recognizes that some makes (or misses) are better than others and captures a player’s ability to execute quality attempts consistently. We demonstrate that command captures underlying skill more effectively than traditional make-or-miss statistics; early-season command predicts late-season success more reliably than traditional shooting percentage. To identify what drives command, we define launch-based metrics assessing consistency in release velocity, angle, and 3D position. Players with greater touch, i.e., more consistent launch dynamics, exhibit stronger command as they can reliably control their shot trajectory. Finally, we develop a physics model to identify the range of launch conditions that result in a make and to determine which launch conditions are most robust to small perturbations. This framework reveals “safe” launch regions and explains why certain players excel at free throws, providing actionable insights for player development.
\end{abstract}

\section{Introduction}
Shooting is widely regarded as the most important skill in basketball. Yet even at the highest levels of the sport, the basketball community still lacks effective metrics capable of characterizing the shooting abilities of individual athletes. Recent advances in sports analytics have demonstrated that player-tracking data can enable more sophisticated and impactful analyses that help players, coaches, and front offices develop richer understandings of performances across the sports world \citep{kovalchik2023player}. 
In 2023, the National Basketball Association (NBA) announced a multi-year partnership with Sony's Hawk-Eye Innovations, which enabled the collection of in-game tracking data \citep{HawkEye23}. The resulting dataset captures high-resolution, 3D pose estimations of players and the ball 60 times per second.
Linking these spatiotemporal data to key events, e.g., shots, passes, rebounds, and fouls, enables new forms of analytics and marks the beginning of a new era in basketball analytics. 

By applying statistical methods to these emerging data sources, we can begin to quantify aspects of player skill that were previously difficult to measure. The goal of this paper is to demonstrate the value of such applications through a detailed exploration of free throw shooting in the NBA. More specifically, we introduce a new ensemble of free throw metrics that provide a richer characterization of an individual shooter's ability to control consistency both at the launch point and at the target location inside the rim itself.

In baseball, \textit{command} refers to a pitcher's ability to locate their pitches \citep{BenPorat2016, StatsPerform2021}.
Previous studies \citep{glanzer2021relationship} have analyzed the relationship between pitcher kinematics and command, showing that differences in mechanics can translate into meaningful variation in pitch placement. We extend the notion of command to free throw shooting in basketball, where a shooter’s ability to effectively control the flight of a shot attempt can similarly distinguish elite performers from average ones, as well as provide more detailed diagnostics about how and why some players struggle to demonstrate consistency at the free throw line.

Traditional shooting indicators such as free throw percentage are coarse, result-oriented metrics that simply summarize the proportion of shots made versus attempted  \citep{kubatko2007starting, terner2021modeling}. While these indicators are simple and interpretable, they reduce each shot to a binary outcome (make or miss), ignoring the flight of the basketball and the underlying launch conditions that actually shape the result. 

As a simple illustration, consider two players who each miss five free throws. The first airballs every attempt, while the second narrowly misses with each shot hitting the rim. Although both players have identical shooting percentages, we intuitively recognize that the second player demonstrates greater shooting skill, as their shots are consistently landing closer to the target location.
We aim to formalize this intuition by defining a command metric in basketball that quantifies not just whether a shot was made but how well it was executed.

Previous studies in basketball have analyzed the relationship between launch characteristics and shooting percentage. Simulation-based work has investigated optimal launch parameters \citep{tran2008optimal}, while others have compared launch characteristics for makes and misses using NBA data (e.g., \citet{maymin2012individual} analyzed 20 players from the 2010-2011 NBA season). Additional research has examined how ball release properties and postural control affect shooting success using data from 25 male college students \citep{verhoeven2016coordination}, explored systematic entry depth and left–right biases and the value of shot data for predicting shooting performance using data from male and female players across all levels \citep{marty2018}, and analyzed ball path curvature and its relationship to performance using 515 free throws from 35 NBA players in the 2023-2024 season \citep{zhu2025}.

In this paper, we present new in-game metrics that advance the richness of free-throw statistics by incorporating 1) ball-launch metrics that determine the nature of the ball's flight at release and 2) the precise landing location of the ball as it reaches the intended target. Our methods investigate the relationships between launch consistency and successful in-rim outcomes. Using a recent sample of NBA free throw shooting that includes over 21,964 attempts by 72 unique NBA players, we combine these metrics with a physics-based model and an optimization algorithm to more precisely flag the specific launch errors that actually cause missed free throws. Finally, our results also allow us to suggest ideal launch specifications for individual shooters in the NBA, enabling new pathways for athletes and teams to improve skill acquisition techniques.  

The paper is organized as follows.
In Section~\ref{sec:data} we describe the NBA data.
Section~\ref{sec:command} defines the new command metric and demonstrates that it is more predictive of future shooting percentage than past shooting percentage.
We define the novel launch metrics in Section~\ref{sec:consistency} and examine their relationship with command. 
In Section~\ref{sec:physics} we construct the physics-based model for free throw shooting, and demonstrate its utility in identifying launch conditions that lead to makes.
We conclude and discuss future work and limitations in Section~\ref{sec:discuss}.

\section{Data}
\label{sec:data}
Our dataset consists of NBA regular season free throws extracted from Hawk-Eye's optical tracking technology, which captures NBA player and ball movements in-game \citep{HawkEye23}. The data records the full flight path of the ball 60 times per second, enabling us to compute launch characteristics (e.g., launch angle, velocity and position) and landing location within the basket. Using the frame-by-frame positional data, we compute the ball’s instantaneous velocity components at each time step. From the initial velocity vector, we obtain the total launch speed $v_0$ and the corresponding launch angle $\theta_0$. Table~\ref{tab:launch-charac} displays the league-wide average and standard deviation for each of the launch characteristics.

\begin{table}[ht]
\centering
\begin{tabular}{lrrr}  
\makecell{} & 
\makecell{Launch Velocity\\ \small MPH} & 
\makecell{Launch Angle\\ \small Degrees} & 
\makecell{Launch Height\\ \small Feet} \\
\hline
League Average & $14.74 \pm 0.33$ & $48.66 \pm 2.99$ & $8.89 \pm 0.42$ \\
\hline
\end{tabular}
\caption{League average launch characteristics. Values shown are launch velocity (MPH), launch angle (degrees), and launch height (feet), with mean and standard deviation.}
\label{tab:launch-charac}
\end{table}

Our dataset originally included $101,679$ free throw attempts across the 2023--2024 and 2024--2025 NBA seasons. However, we found that data from the 2023--2024 season was substantially noisier than data from 2024--2025. Therefore, for the analyses in this paper, we restricted our dataset to the 2024--2025 season, leaving us with $49,562$ attempts. With this dataset, we removed outliers to account for measurement noise and intentional misses (e.g., attempts where the objective is to collect the rebound instead of making the shot). Specifically, we excluded any free throw attempt whose launch position, angle, velocity, or landing location fell more than four standard deviations from the mean. This filtering step helped eliminate intentional misses (i.e., where the ball comes out at a low angle and fast) and erroneous readings from the Hawk-Eye tracking system (i.e., some shot attempts registered at hundreds of miles per hour or negative angles, which were false readings), leaving us with $49,412$ shot attempts.

We restricted our analysis to players with at least 200 free-throw attempts in the 2024--2025 NBA season to ensure reliable estimates of shooting statistics, resulting in $21,964$ shot attempts across 72 of the league's best players. Per our data use agreement, all players are anonymized throughout this paper.

For each free throw attempt, we estimated its precise in-rim landing location, measuring its deviation from the basket's bullseye. 
The basket's bullseye is defined as 2 inches behind the rim's center and represents the optimal location for the ball to enter the rim \citep{marty2018}.
For each qualifying player, we aggregated the mean landing distances of the ball from the basket’s bullseye $\mu$ and the standard deviation of those distances $\sigma$ across all shots. These quantities, which represent the player's inaccuracy $\mu$ and variability $\sigma$, are used to compute the player's command as defined in Equation~\ref{eq:equation1} and described in more detail in the following section. For each shot we also computed the velocity, angle, and 3D position at launch. These values are used to compute the consistency metrics in Section~\ref{sec:consistency} and are used as inputs into the physics-based model in Section~\ref{sec:physics}.

\section{Command}
\label{sec:command}
\begin{figure}[ht!]
    \centering
    \includegraphics[width=0.6\linewidth]{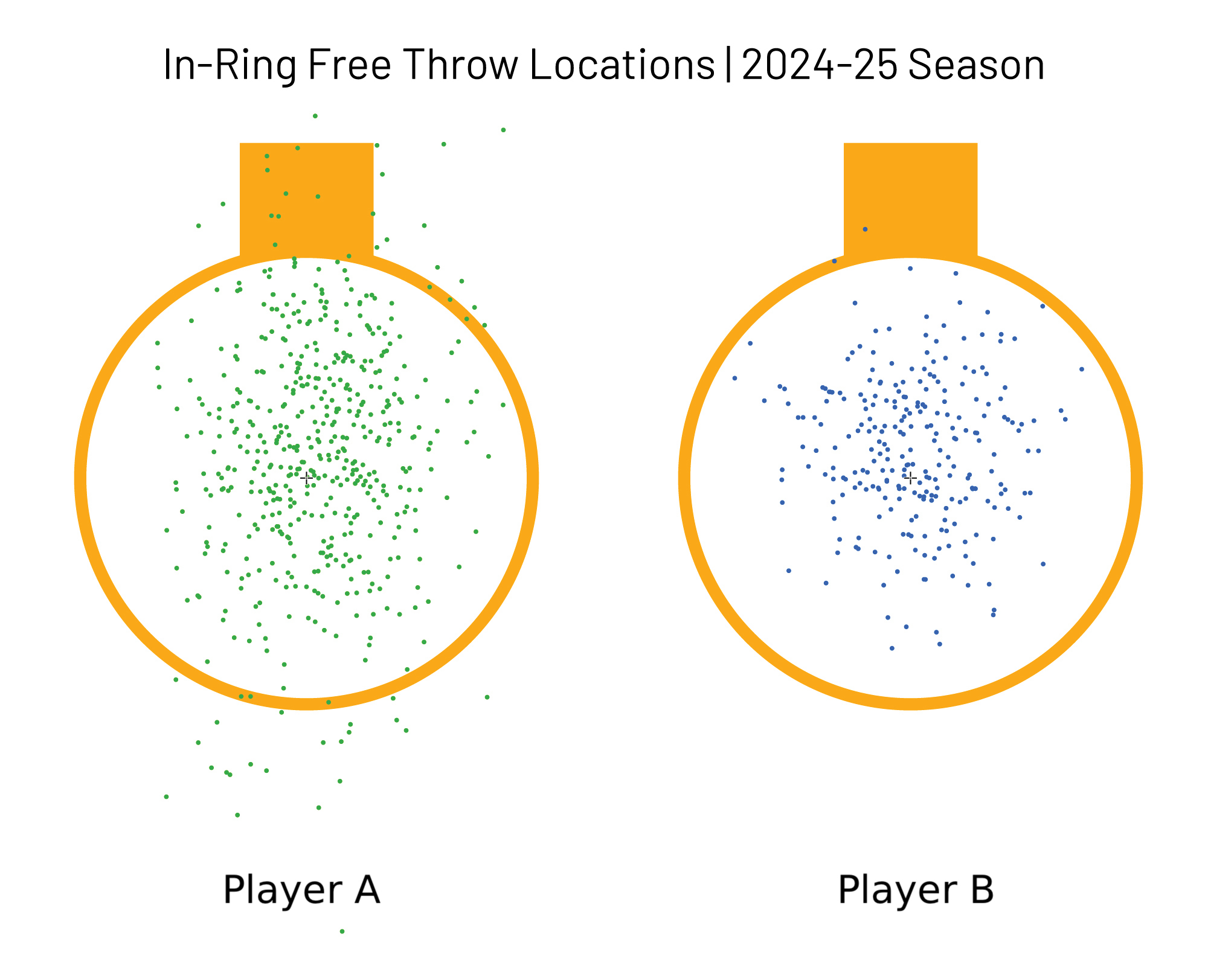}
    \caption{In-ring data for Player A and Player B from the 2024--2025 season. Player B is more accurate and precise than Player A.}
    \label{fig:giannis_steph_spread}
\end{figure}

Using high-resolution spatiotemporal ball-tracking data from the Hawk-Eye system, we examined each free throw attempt’s entry point as it crossed the rim. 
We calculated each player's inaccuracy $\mu$, defined as the ball's average end distance from the bullseye, and variability $\sigma$, defined as the ball's shot-to-shot variation in landing distance from the bullseye.
As an example, Figure~\ref{fig:giannis_steph_spread} visualizes all free-throw attempts by two players from the 2024--2025 NBA season, whom we refer to as Player A and Player B throughout this paper to illustrate key concepts. 
Player B is both accurate and precise: the mean landing point of his shots is close to the bullseye and shows low variability (low $\mu, \sigma$). In contrast, Player A is neither accurate nor precise, i.e., his average landing point is far from the center and his shots exhibit high variability (high $\mu, \sigma$). 

\begin{figure}[ht!]
    \centering
    \includegraphics[width=0.6\linewidth]{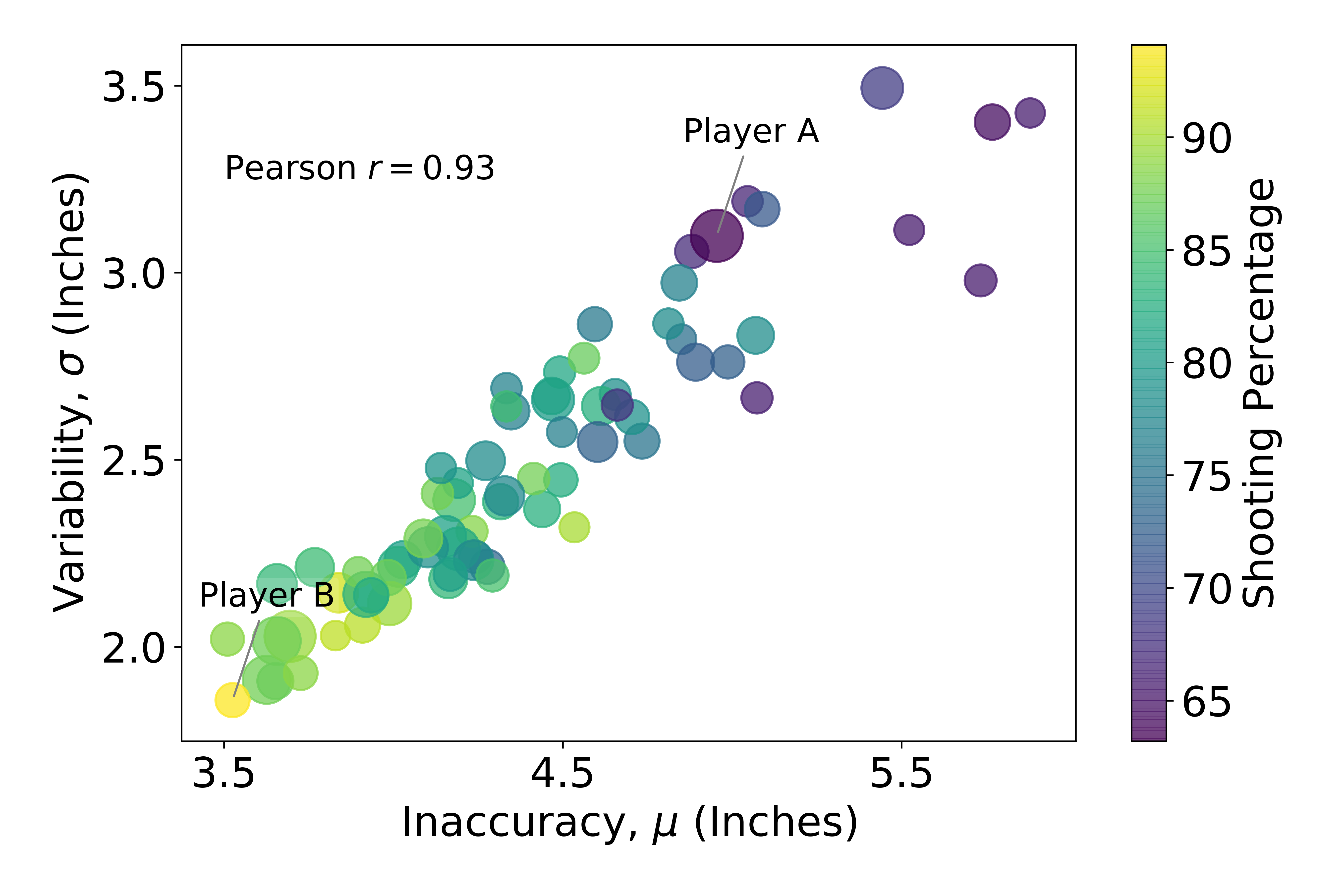}
    \caption{Elite shooters not only hit near the bullseye on average (low mean distance, $\mu$), but do so consistently (low shot-to-shot variability, $\sigma$).}
    \label{fig:league_accuracy_precision}
\end{figure}

Figure \ref{fig:league_accuracy_precision} plots a player's shot variability $\sigma$ versus their shot inaccuracy $\mu$ for all players in our dataset. This scatterplot exhibits a strong positive correlation, indicating that players with greater accuracy tend to also exhibit lower shot variability.

Building on this idea, we introduce the notion of a shooter's \textit{command} inspired by the concept of a pitcher's command from baseball. A commanding shooter is accurate (hits the bullseye on average) and precise (exhibits small deviations in shot dynamics); i.e., they consistently hit the basket's bullseye. This new metric distinguishes players who consistently swish from those who merely `get-lucky' sometimes and roll-in a make. We define a shooter's command to be bounded between 0 and 1 as
\begin{equation}
    C = \frac{1}{1 + \sqrt{\mu^2+\sigma^2}}
    \label{eq:equation1}
\end{equation}

\begin{figure}[ht!]
    \centering
    \includegraphics[width=0.6\linewidth]{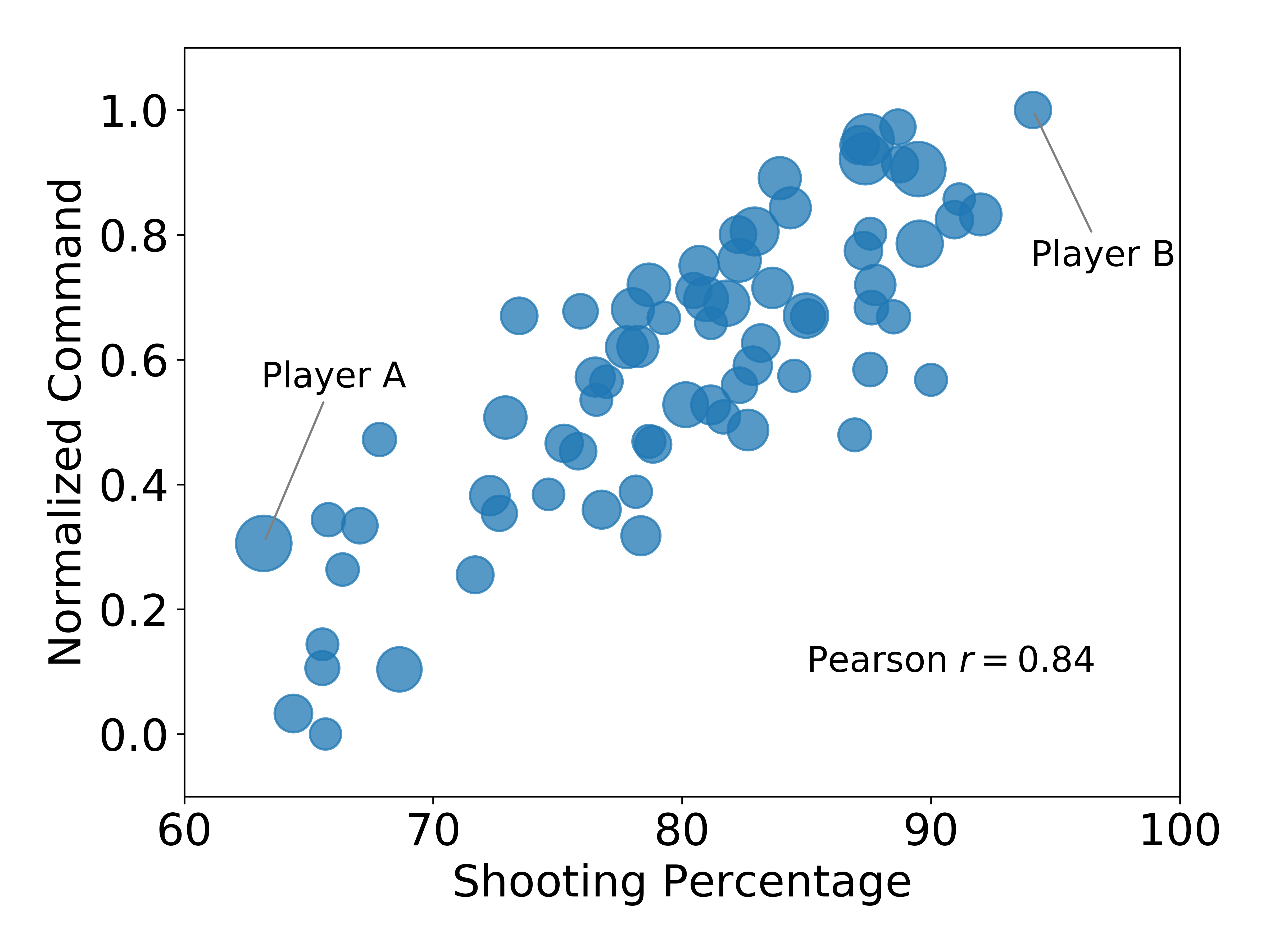}
    \caption{Command correlates with success: players with higher shooting percentages tend to exhibit stronger command --- i.e., both precision and accuracy in their shot. Size of points indicates number of shot attempts.}
    \label{fig:command_percentage}
\end{figure}

Using Equation \ref{eq:equation1} we calculated command for each player. Figure \ref{fig:command_percentage} plots normalized command versus shooting percentage, demonstrating that command exhibits a strong, positive correlation with shooting percentage.

\begin{figure}[ht!]
    \centering
    \includegraphics[width=0.6\linewidth]{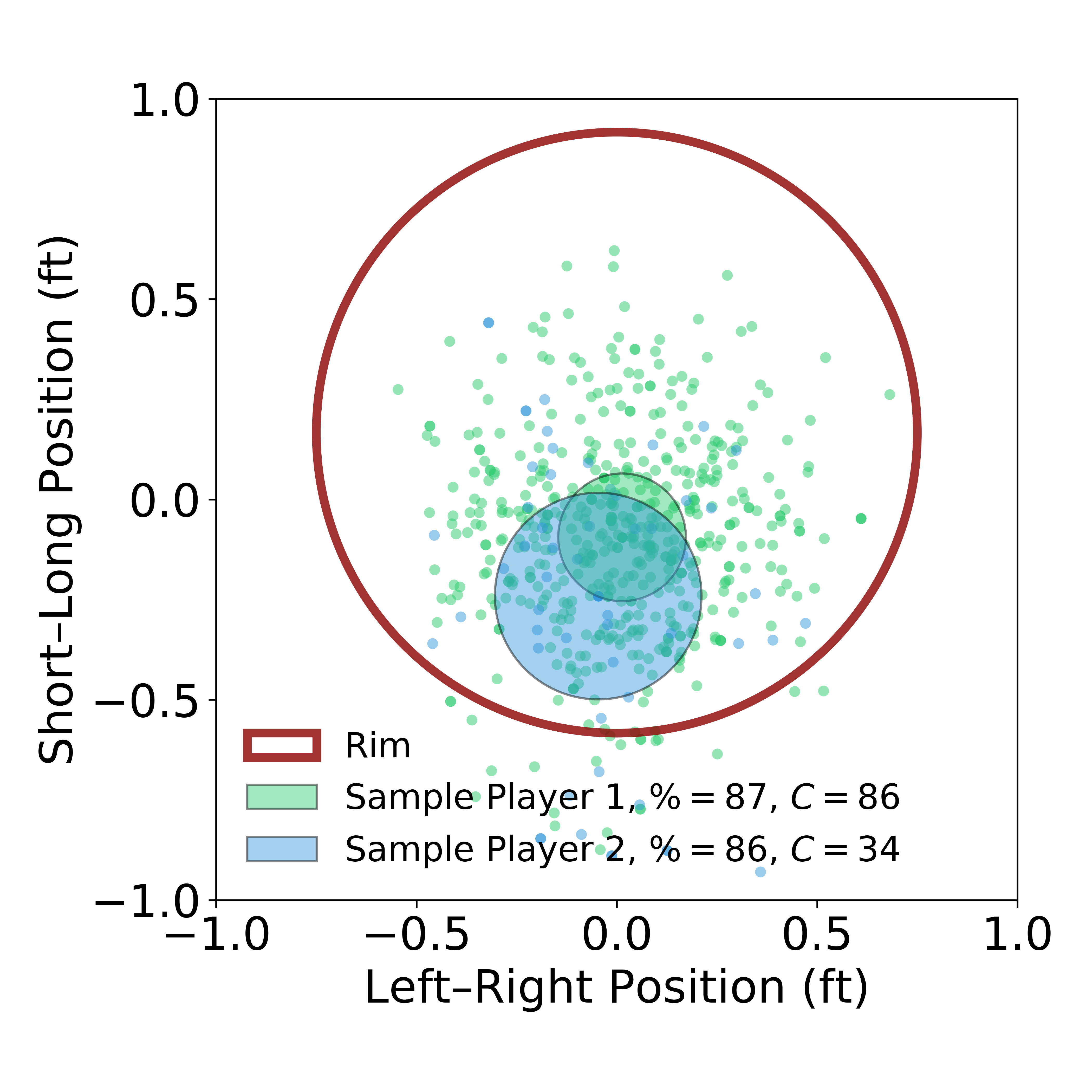}
    \caption{Command captures differences in shooting quality among players with similar shooting percentages. The red circle marks the rim, and each scatter point shows an individual shot landing location for two sample players in our dataset. Large circles indicate each player’s average landing point and spread. More commanding shooters produce smaller, more tightly clustered circles around the bullseye.}
    \label{fig:harden_holmes}
\end{figure}

Command offers a more detailed metric than shooting percentage as it rewards players who are both accurate (on the bullseye, low $\mu$) and precise (with little variation, low $\sigma$). As such, this new metric differentiates between players who post similar shooting percentages but are viewed as different quality shooters. For example, Figure \ref{fig:harden_holmes} illustrates free-throw outcomes for two players that both shot between 86-87\% from the line in the 2024--2025 season, yet placed in the 95th and 48th percentiles of command, respectively, over the same time frame. This indicates that despite comparable shooting percentages, the player in the 95th percentile of command likely exhibits greater reliability and consistency in their performance.

\subsection{Predictive validity of command}

Because command is a more information-dense metric than shooting percentage, we find that it predicts future performance more reliably, especially in small samples. By embedding information about the underlying quality of each shot, command serves as a stronger indicator of long-term success than traditional shooting percentage.

To demonstrate that command is a strong predictor of success in limited sample sizes, we divided the 2024--25 NBA season into “early” and “late” samples, splitting the data at November 15, 2024. We calculated both early and late season shooting percentage and command for all players in the data set with at least 50 shot attempts before and after November 15, 2024. We found that early season shooting percentage predicts late season shooting percentage with a Pearson correlation of $r = 0.61$. However, early season command predicts late season success stronger, with a higher Pearson correlation of $r = 0.67$.
This result highlights that command is a robust predictor of success, particularly in small sample sizes, because it captures more information in shot quality than simple binary shooting percentages. We hypothesize that command could be particularly valuable for NBA scouting, where sample sizes are often limited, because it captures underlying shot quality and provides insight beyond a simple make-or-miss percentage.

\section{Consistency}
\label{sec:consistency}
We previously defined command as an information-rich measure of shot quality that captures a shooter’s accuracy and precision in executing free throws. However, like the shot result itself, command is ultimately an outcome. The ball's trajectory, and thus the result of the free throw, is determined the instant the ball leaves the shooter’s hand.

Therefore, we hypothesize that greater control over the shot's initial conditions, i.e., its launch angle and velocity, correlates with better outcomes (i.e., stronger command). In particular, players who consistently reproduce their preferred launch conditions should achieve more desirable outcomes, whereas those exhibiting greater variability in their shot dynamics should demonstrate lower command. Because a free throw is a closed task without defensive interference, we expect the most commanding shooters to be those who maintain precise and consistent control over their launch parameters. We test these hypotheses by examining how NBA players manage their launch mechanics across repeated free throws, focusing on the consistency of their chosen shot.

To quantify each player’s consistency in launch dynamics, we computed the standard deviation of their launch velocity, launch angle, and 3D launch position relative to the basket. These standard deviations in launch dynamics represent a player's inconsistency in each metric at launch, with larger values indicating greater inconsistency. We define the inconsistency score in launch metric $i$ to be the standard deviation of launch characteristic $i$ (i.e., velocity, angle, or position) for player $p$, $\sigma^{p}_i$. 

We standardized each inconsistency score by converting $\sigma^{p}_i$ into a $z$-score, which preserves relative differences and allows for clearer comparison across metrics:
\begin{equation}
    z^p_i = \frac{\sigma^p_i - \mu_{\sigma_i}}{\sigma_{\sigma_i}}
\end{equation}
where $\mu_{\sigma_i}$ and $\sigma_{\sigma_i}$ are the leaguewide mean and standard deviation of inconsistency in metric $i$. 

We then normalized each $z$-score and define a player’s consistency in metric $i$ as one minus their normalized inconsistency:
\begin{equation}
    r^p_i = 100\% - \text{Normalized}(z^p_i)
\end{equation}
so that higher values of $r^p_i$ indicate more consistent (repeatable) launch behavior (i.e., the player shows lower variability in launch metric $i$).

To obtain an overall consistency measure, we combined the $z$-scores for launch angle $\theta$ and velocity $v$, normalized their sum, and again subtracted from 100\%:
\begin{equation}
    R^p = 100\% - \text{Normalized}(z^p_\theta + z^p_v)
\end{equation}
We excluded deviations in launch position from the overall consistency score as variations in launch velocity and angle can compensate for small positional differences. Thus, $R^p$ reflects how repeatable a player’s launch velocity and angle are across their free throw attempts, with larger values indicating greater consistency. We called a player's overall consistency $R^p$ their \textit{touch}, as players with high touch can reliably control the speed and arc of their shots (corresponding to high consistency in launch dynamics).

\begin{figure}[ht!]
    \centering
    \includegraphics[width=0.7\linewidth]{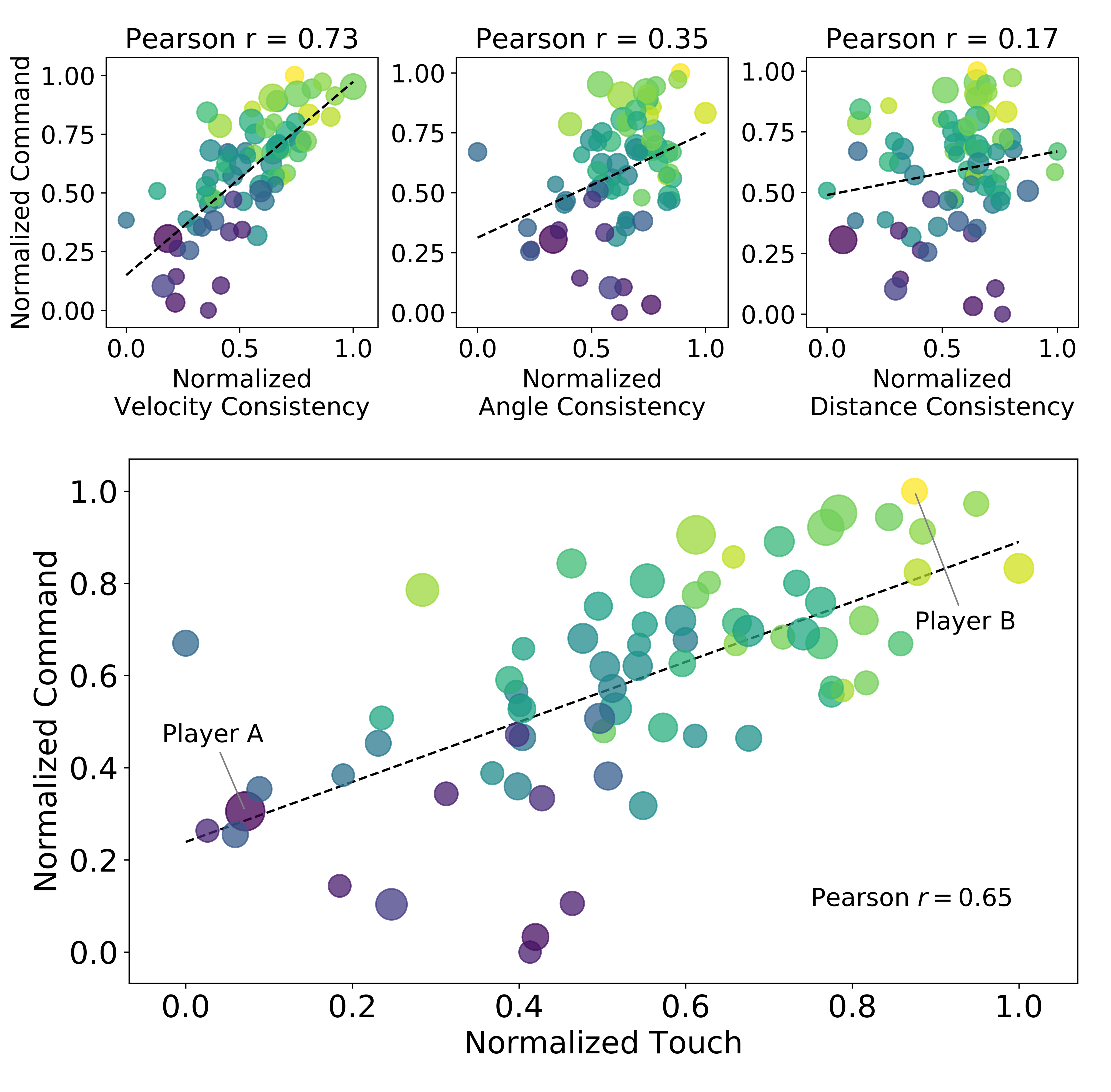}
    \caption{Touch (or, control over launch dynamics) drives a player's command over their shot. Normalized command correlates with normalized touch (or, overall launch consistency) with a Pearson's correlation coefficient $r=0.65$. Point size and color represent the number of attempts and shooting percentage, respectively. Sub-panels show that command also correlates with consistency in individual launch components --- velocity $r=0.73$, angle $r=0.35$, position $r=0.17$ --- with the same visual encoding for attempts and shooting percentage.}
    \label{fig:touch_command}
\end{figure}

Figure \ref{fig:touch_command} highlights that, across the entire league, strong consistency in launch dynamics (or, great touch) contributes to a player's control over their shot, thereby driving high command with a Pearson's correlation coefficient $r=0.65$.
The sub-panels of Figure \ref{fig:touch_command} show how the individual components of a shot's initial launch conditions influence command. We find that consistency in launch position demonstrates the weakest relation with command as variations in velocity and angle adjustments can offset positional errors. In contrast, consistency in launch velocity is the most critical factor for determining a shooter's command. Launch velocity governs the total energy imparted to the ball, and because both the ball's energy and final landing location scale with the square of velocity, small deviations in launch velocity compound through the physics mapping to produce large variations in outcome (detailed in Section~\ref{sec:physics}). Consistency in launch angle also contributes to strong command, however, it is less influential than velocity. In the physics mapping, launch angle affects the ball's landing location through a sin$\theta$ $\cdot$ cos$\theta$ term which is locally flat around $\theta \sim 45 ^\circ$, the typical release angle for most players. Therefore, small deviations in launch angle $\theta$ are largely masked by this locally flat slope of the sin$\theta$ $\cdot$ cos$\theta$ term.

\begin{figure}[ht!]
    \centering
    \includegraphics[width=0.9\linewidth]{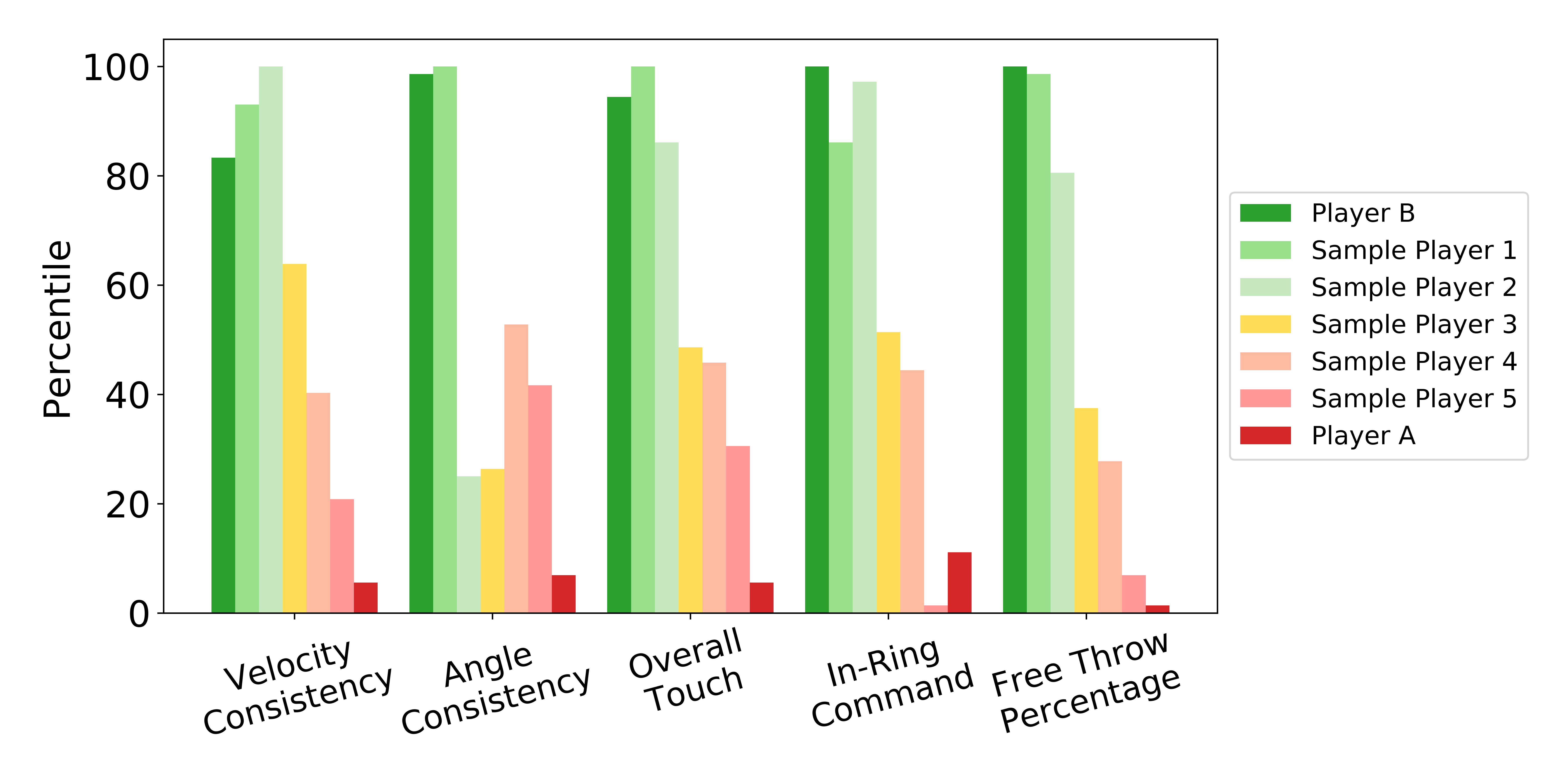}
    \caption{Player-specific case study illustrating how launch inputs influence shot outcomes. Players who exhibit high consistency and precise control over the physics of their shot generally achieve greater success. These players are often smaller, more skilled guards or wings, whereas players with lower consistency tend to be larger, less agile bigs.}
    \label{fig:touch_command_casestudy}
\end{figure}

Next, to further illustrate that touch (launch consistency) drives strong command, we examine seven specific players who span roughly evenly spaced percentiles across the league in touch, command, and free throw percentage. In this case study shown in Figure \ref{fig:touch_command_casestudy}, we observe a staircase effect: as a player’s consistency in launch angle, velocity, and overall touch decreases, so too does the quality of their shot outcomes, as measured by command and shooting percentage.

Finally, in Table \ref{tab:table1} we converted the consistency, touch, command, and free throw percentage statistics to percentiles for easy comparison. This table presents the ten players with the highest and lowest command in the 2024--2025 NBA season. Players with high touch and strong command are typically smaller, skilled guards and wings, whereas larger, less agile bigs generally show lower command and weak control of their launch (weak touch). The table also highlights that players may struggle with different aspects of their launch dynamics. For instance, Player 2 ranks in the 74th percentile for consistency in launch angle but only the 7th percentile for launch velocity consistency. Given that launch velocity sensitively alters the ball's trajectory, these results suggest that Player 2 could improve his shot outcomes by focusing on controlling the velocity of his releases and, thus, his overall touch.

\begin{table}[ht!]
\centering
\resizebox{0.9\textwidth}{!}{  
\begin{tabular}{
    L{4.5cm}   
    R{1.7cm}   
    R{2.15cm}   
    R{2.15cm}   
    R{2.15cm}   
    R{2.15cm}   
    R{2.15cm}   
}
 &  & \multicolumn{3}{c}{Launch} & \multicolumn{2}{c}{Response} \\
\cmidrule(lr){3-5} \cmidrule(lr){6-7}
\makecell{Player} & 
\makecell{Attempts} & 
\makecell{Velocity\\Consistency\\Percentile} & 
\makecell{Angle\\Consistency\\Percentile} & 
\makecell{Overall\\Touch\\Percentile} & 
\makecell{Command\\Percentile} & 
\makecell{Free Throw\\Percentage\\Percentile} \\
\midrule

Player 1 & 201 & 21 & 42 & 31 & 1 & 7 \\
Player 2 & 292 & 7 & 74 & 32 & 3 & 3 \\
Player 3 & 405 & 4 & 32 & 14 & 4 & 14 \\
Player 4 & 238 & 32 & 46 & 36 & 6 & 4 \\
Player 5 & 209 & 8 & 15 & 8 & 7 & 6 \\
Player 6 & 279 & 12 & 4 & 4 & 8 & 15 \\
Player 7 & 217 & 10 & 6 & 3 & 10 & 10 \\
Player 8 & 633 & 6 & 7 & 6 & 11 & 1 \\
Player 9 & 314 & 56 & 38 & 51 & 12 & 39 \\
Player 10 & 261 & 38 & 31 & 33 & 14 & 11 \\
$\vdots$ & $\vdots$ & $\vdots$ & $\vdots$ & $\vdots$ & $\vdots$ & $\vdots$ \\
Player 62 & 362 & 93 & 100 & 100 & 86 & 99 \\
Player 63 & 345 & 19 & 56 & 35 & 88 & 69 \\
Player 64 & 203 & 51 & 78 & 67 & 89 & 97 \\
Player 65 & 367 & 72 & 71 & 74 & 90 & 68 \\
Player 66 & 609 & 65 & 43 & 64 & 92 & 92 \\
Player 67 & 267 & 99 & 69 & 97 & 93 & 90 \\
Player 68 & 538 & 86 & 68 & 82 & 94 & 79 \\
Player 69 & 303 & 94 & 81 & 92 & 96 & 76 \\
Player 70 & 535 & 100 & 25 & 86 & 97 & 81 \\
Player 71 & 256 & 96 & 97 & 99 & 99 & 89 \\
Player 72 & 271 & 83 & 99 & 94 & 100 & 100 \\

\bottomrule
\end{tabular}
}
\caption{Touch (or, launch consistency) shapes shot outcomes. Overall touch is decomposed into angle and velocity consistency (percentiles). The table highlights ten players with the highest and lowest command (percentiles) from the 2024--2025 NBA season, illustrating that command is driven by consistent launch dynamics.}
\label{tab:table1}
\end{table}

\section{Physics-based model}
\label{sec:physics}
To relate a particular shot's launch conditions to its outcome, we constructed a physics-based model of a basketball free throw. For simplicity, we considered a two-dimensional trajectory neglecting any left-right drift. We also ignored the effects of air resistance (drag) and lift generated by spin (Magnus effect) as a quick estimate shows these forces are small (less than $6\%$ of the ball's weight) for a typical shot with velocity $v = 12$ MPH and spin rate $\omega = 2$ rev/s. As drag and lift forces are negligible and omitted from the model, the ball's trajectory can be described as simple projectile motion.

\begin{figure}[ht!]
    \centering
    \includegraphics[width=0.9\linewidth]{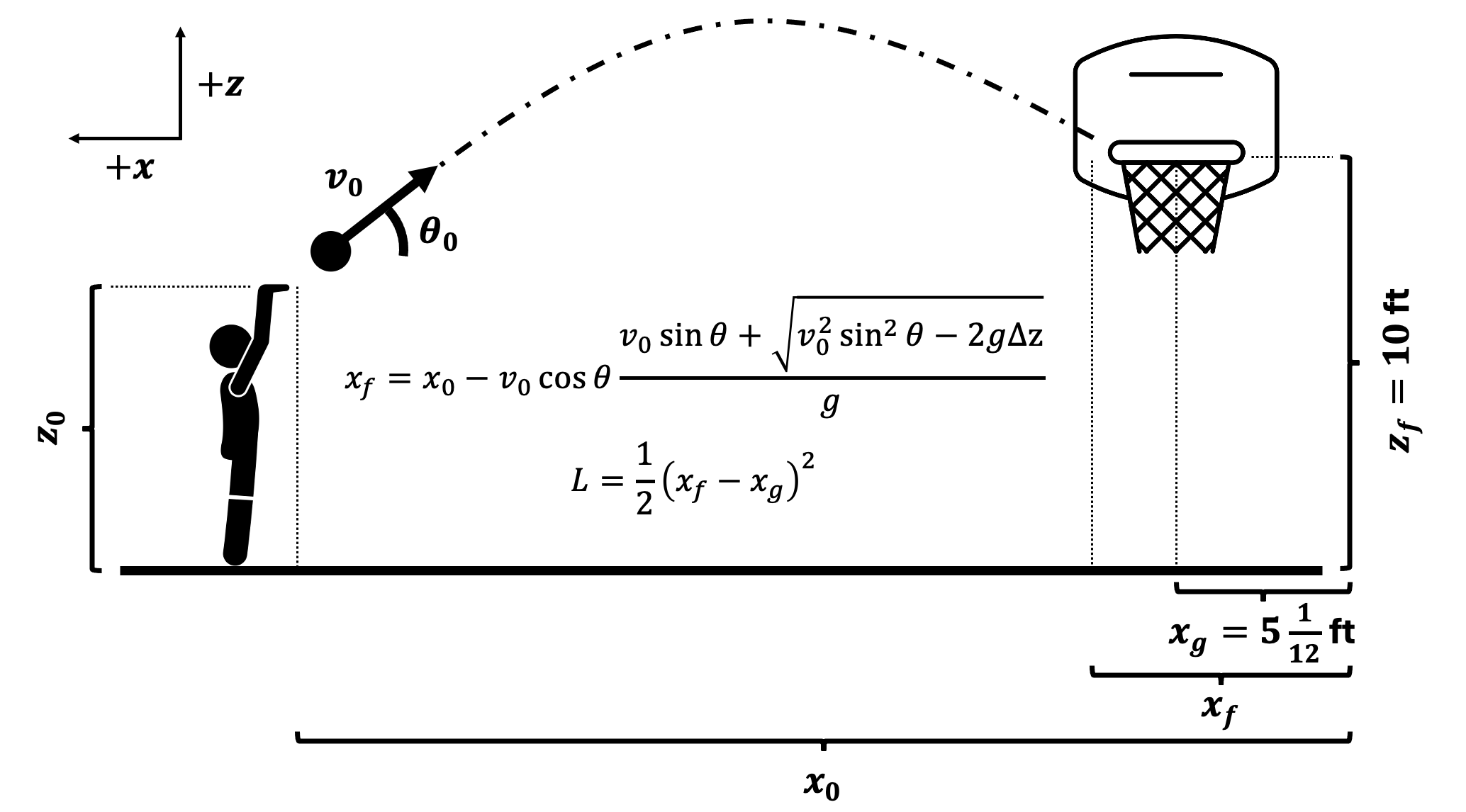}
    \caption{Physics model schematic. A player launches the ball with some initial velocity $v_0$ at an angle $\theta_0$ from some position $x_0, z_0$. The trajectory of the ball travels a horizontal distance $x_f$ once the ball reaches its target location at the rim $z_f=10$ ft.}
    \label{fig:physics_model}
\end{figure}

To model the shot trajectory, we defined the initial launch conditions as the ball’s position $x_0,z_0$ (where $x_0$ is the horizontal distance to the baseline and $z_0$ is the height above the floor), launch angle $\theta_0$, and launch velocity $v_0$. The horizontal position of the ball when it reaches the rim $x_f$ is given by
\begin{equation}
    x_f = x_0 - v_{0x} \Delta t
\end{equation}
where $v_{0x} = v_{0} \text{cos}\theta$ is the horizontal component of the launch velocity and $\Delta t$ is the flight time before the ball reaches the rim. The flight time can be determined from the ball's vertical motion
\begin{equation}
    \Delta z = v_{0z} \Delta t - \frac{1}{2} g \Delta t^2
\end{equation}
where $v_{0z} = v_{0} \text{sin}\theta$ is the vertical component of the launch velocity, $\Delta z = z_f - z_0$ is the change in height from release to rim, and $g$ is the gravitational acceleration. Solving for $\Delta t$ and substituting back into the horizontal equation yields the horizontal landing position of the ball $x_f$ when it reaches the rim height $z_f=10$ ft as a function of the release position, velocity and angle:
\begin{equation}
    \label{eq:equation6}
    x_f = x_0 - v_{0} \text{cos}\theta \frac{\text{sin}\theta + \sqrt{v_0^2 \text{sin}^2\theta - 2g\Delta z}}{g}.
\end{equation}
A schematic of the two-dimensional projectile model is shown in Figure \ref{fig:physics_model}.

\begin{figure}[ht!]
    \centering
    \includegraphics[width=0.9\linewidth]{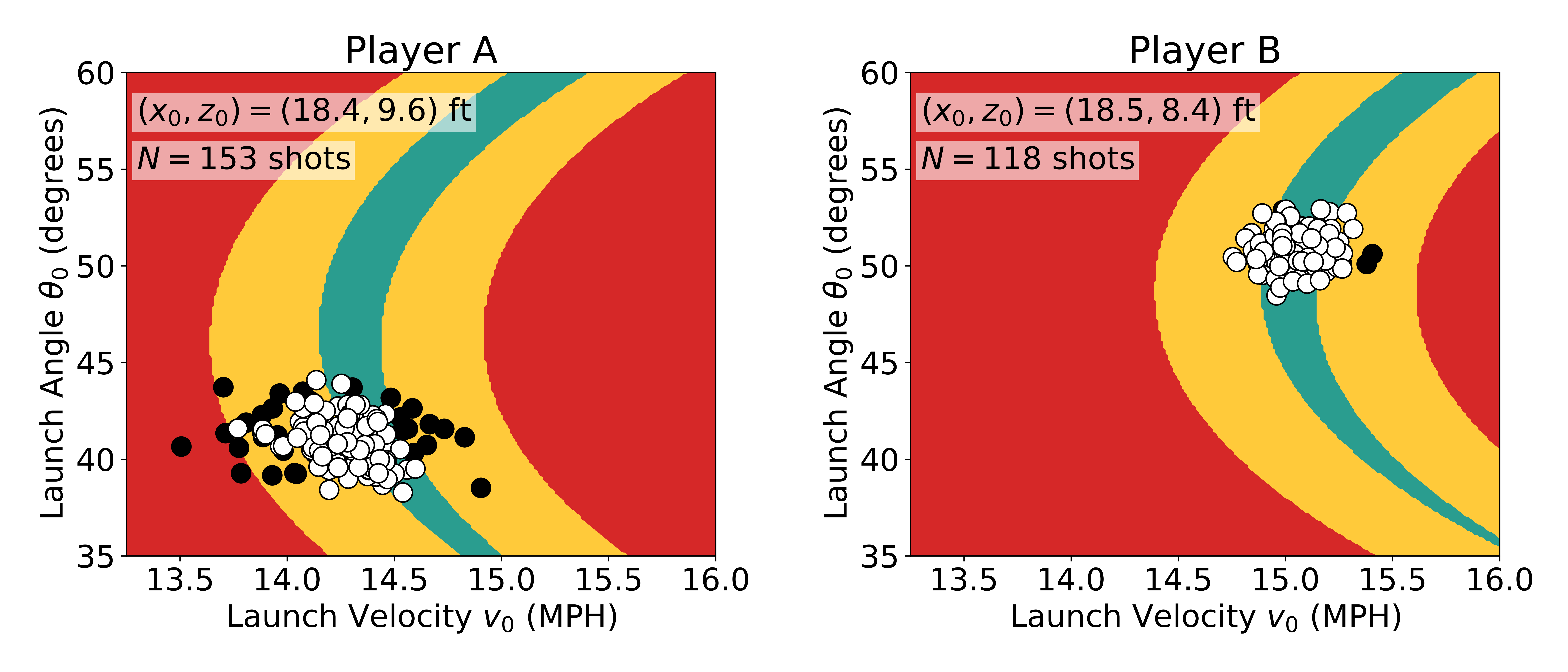}
    \caption{Launch conditions $v_0, \theta_0$ that guarantee success for Player A and Player B. The physics model identifies a band of angle-velocity combinations that result in a swish (green), hit the rim (yellow), or miss entirely (red). Points overlayed represent misses (black) and makes (white) from the Hawk-Eye dataset.}
    \label{fig:launch_conditions_giannis_steph}
\end{figure}

The model in Equation \ref{eq:equation6} depends on the shot’s initial position $(x_0, z_0)$, launch velocity $v_0$, and launch angle $\theta_0$. Using this model, we can determine the shot outcome for any combination of $v_0$ and $\theta_0$ at a fixed launch location $(x_0, z_0)$. In Figure \ref{fig:launch_conditions_giannis_steph}, we evaluated the model over a range of launch velocities and angles to map out the resulting shot outcomes for typical release positions of Player A at $(x_0, z_0) = (18.4,9.6)$ ft and Player B at $(x_0, z_0) = (18.5,8.4)$ ft.

The model reveals a band of velocity-angle combinations that result in a perfect swish (the green region in Figure \ref{fig:launch_conditions_giannis_steph}). As the launch angle deviates towards more extreme values ($>55^\circ$ or $<40^\circ$), this optimal region narrows, indicating reduced tolerance to variation in launch speed. Around $\sim 45-50^\circ$, the width of the green optimal band reaches its maximum. The model also predicts shots that contact the rim, either front or back, shown in yellow. These represent marginal cases where the shot outcome is uncertain and lies beyond the resolution of our simplified model. For more extreme launch conditions, the model identifies shots that completely miss the rim (red region). Finally, we overlay empirical data from the Hawk-Eye dataset which illustrate that actual made shots cluster within the model’s predicted optimal region. Also, because Player B releases the ball more than a foot lower than Player A, the model predicts that his optimal shot requires a higher velocity and a steeper launch angle.

\subsection{Identifying error suppressing launch conditions}
One of the central ideas of this paper is that commanding the foul shot requires touch, that is, precise control, repeatability, and consistency in the launch parameters at the line: controlling $\theta_0$ and $v_0$ is critical. Using this physics-based model, we can examine how small deviations in these launch parameters $\theta_0, v_0$ propagate through the dynamics to produce variations in the shot’s landing position $x_f$ and, ultimately, shot outcomes.

For example, consider the phase space of possible launch conditions for Player A in Figure~\ref{fig:launch_conditions_giannis_steph}. If Player A releases the ball with the combination ($v_0, \theta_0 = 14.4~\text{MPH}, 46^\circ$), the attempt results in a perfect swish. Now imagine he loses control at launch and the ball comes off slightly flatter, say at $44^\circ$; in this case, the shot still swishes. However, if the original launch combination were ($v_0, \theta_0 = 14.5~\text{MPH}, 39^\circ$), and the same small angular error shifted the release to $37^\circ$, the shot would no longer be guaranteed to go in. This example illustrates that some launch combinations are inherently safer, i.e., more robust to small errors, than others.

To quantify this idea, we examine every launch combination for Player A and Player B in Figure~\ref{fig:launch_conditions_giannis_steph}. For each pair $(v_0, \theta_0)$, we assume that the release is subject to a small, fixed perturbation such that $v_0 \pm \delta_{v_0}$ and $\theta_0 \pm \delta_{\theta_0}$. For simplicity, we take these error scales $\delta_{v_0}$ and $\delta_{\theta_0}$ to be the empirical standard deviations of each player’s launch velocities and launch angles, reflecting the inherent variability in their shooting. For Player B, these small errors at the launch were calculated to be  $\delta_{\theta_0} = 1.11^\circ$ and $\delta_{v_0} = 0.24 \text{ MPH}$, and for Player A, $\delta_{\theta_0} = 1.74^\circ$ and $\delta_{v_0} = 0.28 \text{ MPH}$. We then compute how these perturbations shift the ball’s final landing position $x_f$ relative to the unperturbed shot. In doing so, we can identify which launch conditions amplify small release errors and which are comparatively robust.

\begin{figure}[ht!]
    \centering
    \includegraphics[width=0.9\linewidth]{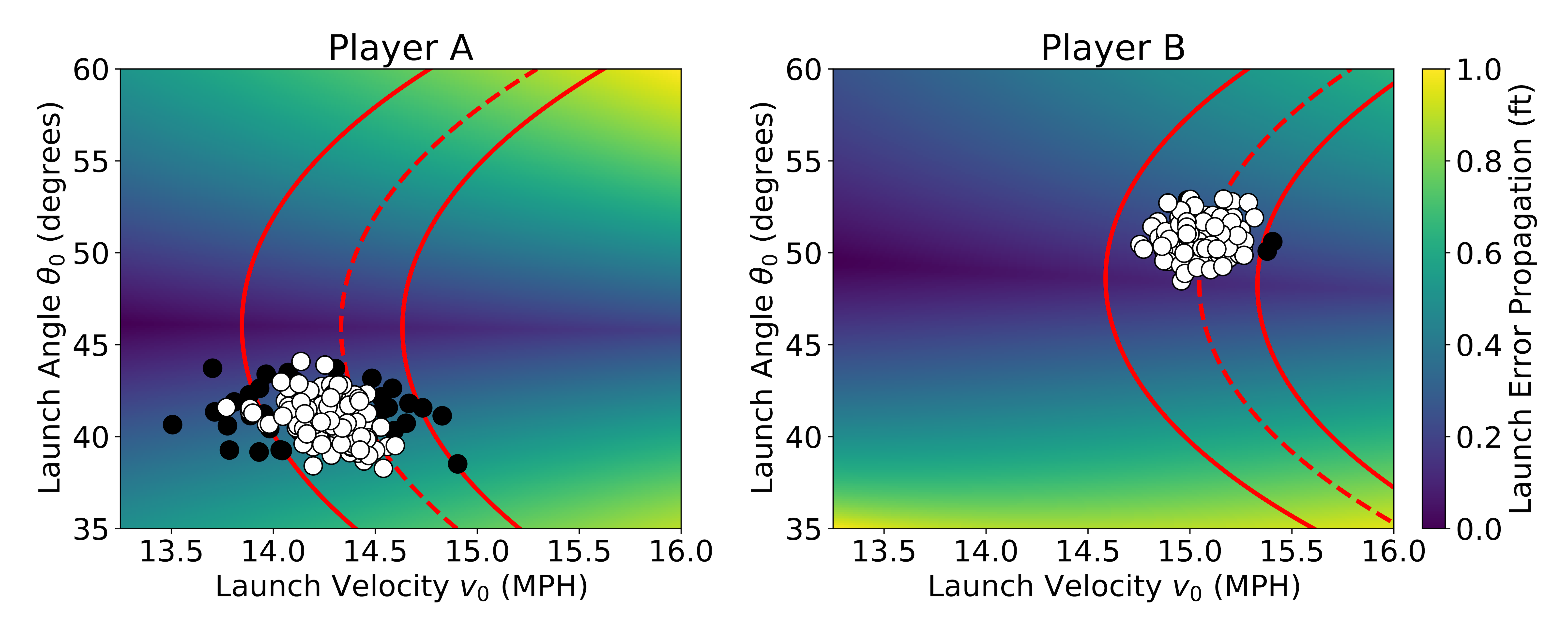}
    \caption{Player B suppresses launch errors more effectively than Player A. For both players, each launch condition $(v_0, \theta_0)$ is perturbed by small errors in velocity and angle, which propagate through the shot dynamics. Dark regions indicate launch conditions that suppress error propagation, while light regions amplify it. Player B’s typical launch (scatter points) lies in a region robust to perturbations, whereas Player A’s does not. White dots denote made shots and black dots denote misses. The red dashed line represents launches that hit the bullseye exactly, and the red solid lines indicate combinations that strike the front or back rim exactly.}
    \label{fig:shot_spread_giannis_steph}
\end{figure}

In Figure~\ref{fig:shot_spread_giannis_steph}, we show how these small, fixed perturbations in $v_0$ and $\theta_0$ shift the ball’s final landing position $x_f$ relative to the unperturbed shot, a quantity we refer to as launch error propagation. Some launch combinations amplify small release errors more than others, and Figure~\ref{fig:shot_spread_giannis_steph} visualizes this sensitivity: yellow regions indicate launch conditions that magnify perturbations and substantially shift the final landing location $x_f$ away from the expected trajectory, while blue regions indicate launch conditions that suppress perturbations and are therefore more robust to inconsistencies at the launch. Notably, the launch errors $\delta_{v_0}$ and $\delta_{\theta_0}$are quite small for both players ($<2^\circ$ and $\sim1/4$ MPH), yet manifest in trajectories that deviate by up to 1 ft in distance (which could be the difference between a swish and an air-ball).

By introducing small perturbations $\delta_{v_0}$ and $\delta_{\theta_0}$ into all possible launch conditions $(v_0, \theta_0)$ for Player A and Player B, we find that some launch combinations are less sensitive to noise than others. In particular, at each player’s typical launch location, we observe a dark band around $46^\circ$ for Player A and $50^\circ$ for Player B. These dark bands correspond to launch conditions that suppress the propagation of small errors, whereas higher or lower angles amplify error. Additionally, for both players, increasing velocity amplifies error propagation.

Notably, Player B’s typical launch parameters fall within the dark band of the error-suppression region, whereas Player A's do not. This difference is reflected in Player B’s higher touch and command compared with Player A. For Player A to reduce inconsistencies at the launch and gain touch, these results recommend that he increase the angle of his shot.

From the physics model in Equation~\ref{eq:equation6}, the ball’s final landing location $x_f$ depends on the square of the initial velocity, $v_0^2$, while the launch angle enters through the cross term $\sin\theta \cos\theta$. Around $45^\circ$, the slope of this cross term is locally flat, minimizing the impact of small deviations in $\theta$. In contrast, because $x_f$ depends on $v_0^2$, small deviations in launch velocity are increasingly amplified at higher velocities. Taken together, these effects suggest that to minimize error propagation and reduce sensitivity to noise at the launch, both Player A and Player B should adopt launch characteristics with low velocities and intermediate angles. The optimal shot would therefore be where the error suppressing region in Figure~\ref{fig:shot_spread_giannis_steph} (dark blue band) intersects the bullseye line (red dashed line), coincidentally, exactly where Player B places his shots. Adopting such launch characteristics would guarantee a perfect swish while making each shooter more robust to slight potential errors at the line.

\subsection{Optimizing shots via the physics model}
\begin{figure}[ht!]
    \centering
    \includegraphics[width=0.8\linewidth]{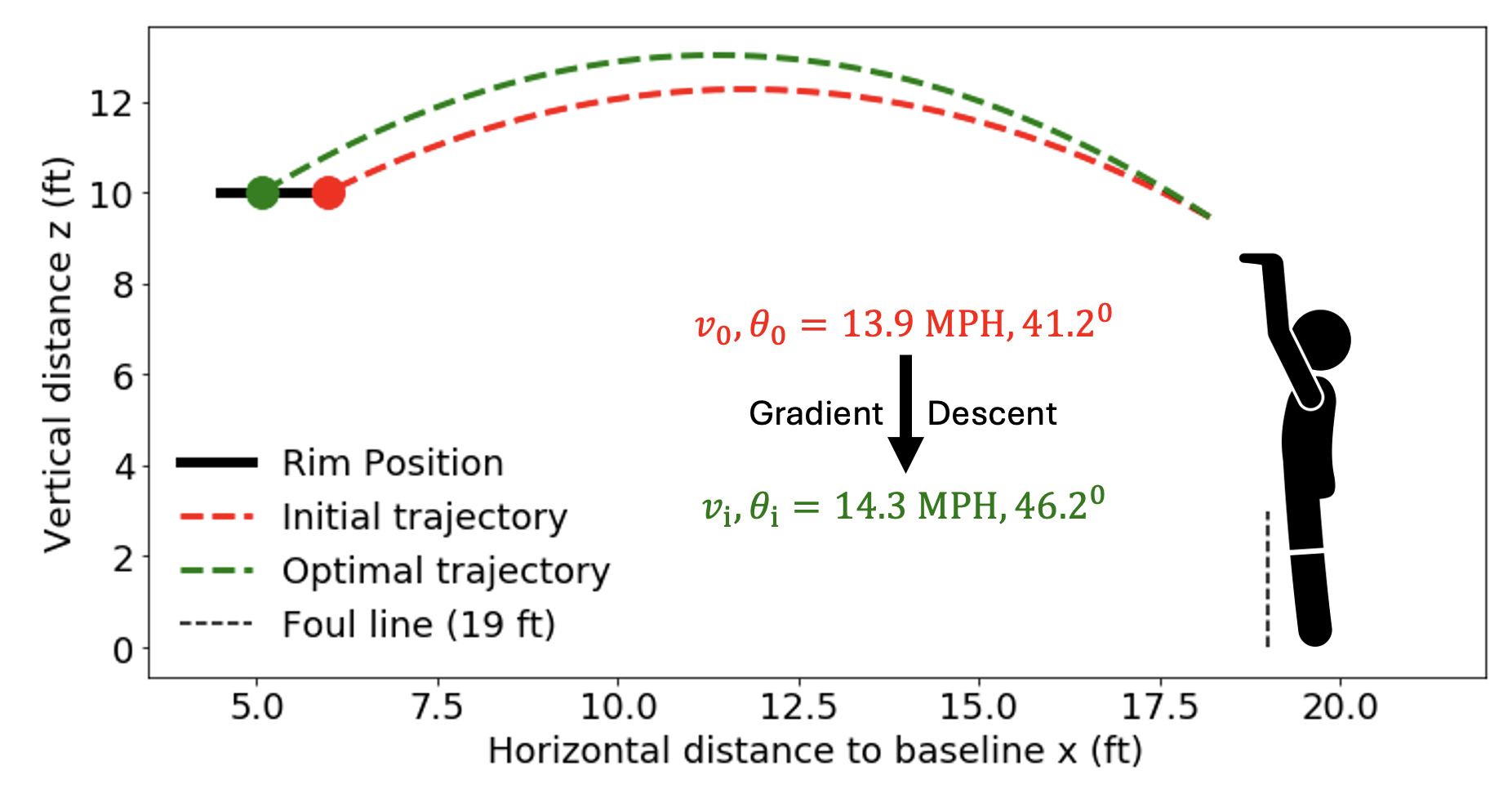}
    \caption{Optimizing missed shots using the physics model. Starting from an initial shot with launch conditions $v_0, \theta_0$ corresponding to the red trajectory, we perform gradient descent on the loss function to identify an optimal shot with launch conditions $v_i, \theta_i$ shown in green. This approach could provide real-time feedback to players following a missed shot.}
    \label{fig:shot_feedback}
\end{figure}

Because the physics model predicts the ball’s horizontal position $x_f$ at the rim height $z_f$ for some initial launch conditions $v_0, \theta_0$, it has the potential to provide real-time feedback to players during practice or games. Assuming the player aims for the basket's bullseye at $x_g = 5 \frac{1}{12}$ ft from the baseline, we define a simple loss function that quantifies the deviation between the desired landing position $x_g$ and the actual landing location predicted by the model $x_f$:
\begin{equation}
    L = \frac{1}{2} (x_f - x_g)^2
\end{equation}
By performing gradient descent on this loss function by iteratively adjusting the initial launch conditions $v_0, \theta_0$ in the direction that reduces the loss, we can identify the optimal shot parameters $v_i, \theta_i$ that minimize the deviation in the ball's landing location $x_f$ from the basket’s bullseye. This procedure determines the launch conditions that produce a perfect swish as demonstrated in Figure \ref{fig:shot_feedback}. This method could be used to provide players with real-time feedback on how to adjust their shots to achieve an optimal trajectory.

\section{Discussion}
\label{sec:discuss}
Our results demonstrate that Hawk-Eye tracking data enable a new way to quantify basketball shooting skill. We introduce the new concept of command in basketball, which is a measure of a shooter's ability to control both the accuracy and precision of the ball's landing location, moving beyond traditional binary metrics such as free throw percentage. The strong correlation between command and future shooting percentage demonstrates that command represents a more fundamental shooting skill. We further demonstrate that consistency in launch characteristics (i.e., launch velocity, angle, and distance), or touch, is strongly associated with command, highlighting the importance of controlling the ball's dynamics. Finally, we present a physics-based model that identifies the optimal launch conditions for a given player, providing an interpretable link between mechanics and shooting outcomes.

While these results are promising, several limitations warrant discussion. First, the command metric is inherently sensitive to noise in the tracking data. We observed a stronger signal in the 2024 data compared to the 2023 data, which may reflect improvements or recalibration in the Hawk-Eye system. Because command is subject to measurement noise, unlike shooting percentage, this noise can bias the command score in very small sample sizes. In general, we find reliable measurements of command in sample sizes larger than $N \approx 50$. As tracking systems continue to improve, estimates of command will become increasingly stable and useful, particularly in contexts where data are limited, such as drafting, scouting, or early-season trades.

In addition, our physics-based model does not explicitly account for certain effects such as drag and Magnus effect as their influence on the trajectory is small ($<6\%$ in typical situations). These forces are likely to contribute relatively minor variation compared to the ball’s inertia, but their inclusion could refine estimates of optimal launch conditions.  For the same reason, we model the ball trajectory in two dimensions; free throws exhibit minimal lateral deviation, so a 2D representation captures the primary mechanics of the shot. A full three-dimensional model may offer additional nuance.

While our analysis focuses on free throws due to their controlled setting, extending the concepts presented here to field goals is an important next step. Unlike free throws, in-game field goals are influenced by additional factors such as defensive pressure and shot selection, which introduce more variability in launch conditions and context. Quantifying command and touch in this setting would provide insight into how players maintain accuracy under more dynamic conditions.

Future research should focus on linking launch characteristics to player biomechanics. Integrating optical tracking with biomechanical data would help explain how physical movements translate into consistent release dynamics and ultimately into higher command. This line of research would provide a better understanding of the mechanical foundations of elite shooting and open opportunities for individualized player development and injury prevention.

\bibliographystyle{agsm}
\bibliography{bib}

\end{document}